\documentstyle[preprint,aps]{revtex}

\begin{document}
\title{{\large {\bf A possible role of $D^-$ band in hopping conductivity
and metal-insulator transition in 2D structures}}} 
\author{{V.I.Kozub and N. V.
Agrinskaya}} 
\address{A. F. Ioffe Institute, Solid State
Physics Division, 194021 Saint Petersburg, Russia } 
\author{{S. I.
Khondaker }} 
\address{Physics Department, Dhaka University, Dhaka 1000,
Bangladesh } 
\author{{I. Shlimak}} 
\address{Jack and Pearl Resnick
Institute of Advanced Technology, Department of Physics,\\ Bar-Ilan
University, Ramat-Gan 52900, Israel} 
\maketitle 
\date{\null}

\begin{abstract}
\baselineskip=3.5ex 
A simple two-band model is suggested explaining
recently reported unusual features for hopping magnetoresistance and the
metal-insulator transition in 2D structures. The model implies that the
conductivity is dominated by the upper Hubbard band ($D^{-}$-band).
Experimental studies of hopping magnetoresistance for Si $\delta $-doped
GaAs/AlGaAs heterostructure give additional evidences for the model.
\end{abstract}

\textwidth=6.0in

\bigskip \baselineskip=3.5ex

\section{Introduction}

In recent papers \cite{Agrin,Polyan}, the existing data on the hopping
magnetoresistance for the nearest neighbor hopping were reconsidered with
a conclusion that in many cases the magnetoresistance is related to a
suppression of a contribution of the upper Hubbard band due to a spin
alignment. A signature of such a behavior is a universal magnetic field
behavior of the activation energy related to the Zeeman splitting. It was
the surprisingly good agreement between the $g$-factor values extracted
from the experimental data with corresponding handbook values which lead
to the conclusion given above.

While this conclusion emphasizing a role of the upper Hubbard band or
$D^{-}$ band was formulated for the nearest neighbor hopping (where it
manifests itself as the Zeeman addition to the activation energy), a
possible contribution of $D^{-}$ band in the variable range hopping
regime was studied theoretically first by Kurobe {\it et al}.
\cite{Kamimur}. It was shown that a presence of $D^{-}$ band at the Fermi
level leads to a suppression of some hopping channels by the magnetic
field and manifests itself as a plateau-type feature in the
magnetoresistance curve. Recently a presence of corresponding feature for
doped CdTe crystals was observed \cite {Agrin2}.

However, the picture considered in \cite{Kamimur} implies that without
external magnetic field, the contribution of the upper Hubbard band to
the conductivity is smaller or at least comparable with that one of the
lower Hubbard band. In what follows we are going to analyze a different
situation assuming that at zero magnetic field the contribution of the
$D^{-}$ band dominates. This assumption is not an exotic one since the
localization length in the $D^{-}$ band is known to be much larger than
in the standard impurity band; so the contribution of the $D^{-}$ band
can dominate even if the corresponding density of states is much less
than for the standard band.

We will assume that a finite density of states related to the upper
Hubbard band ($D^-$ band) exists at the Fermi level; moreover, we assume
that it is this band that mainly contribute to the current because of the
larger localization length. The lower Hubbard band, although having a
larger density of states, is supposed to give much less contribution to
the current and - probably - has localized states at the Fermi level.

We will show that such a simple two-band model can consistently explain
the unusual crossover from variable range hopping conductivity to
activated hopping in strong magnetic field reported recently for 2D
electron layers in \cite{Shlimak}. Moreover, it reproduces the main
features of temperature and magnetic field behavior of the 2D
metal-insulator transition extensively studied in recent years (see
e.g.\cite{Krav,Sar,heter}).

To make a deeper insight into the experimental situation of
\cite{Shlimak} we present here the results of measurements of the
magnetic field dependence of the activation energy of resistivity and its
variation with the gate voltage.

\section{The model}

Note that, although a presence of $D^-$ channel was never disputed,
during last years the hopping conductivity was typically considered as
related to "standard" hopping within lower impurity band \cite{Efros}.
This point of view was, on the one hand, related to the fact that the
only temperature dependence of conductivity does not allow to
discriminate clearly between the contributions of different Hubbard bands
and to estimate the Hubbard energy from experimental data. On the other
hand, the theoretical estimates of the Hubbard energy seem to be too
large to allow $D^-$ band contribution. Thus it seemed that any effect of
the upper band can exist only for nearly completely occupied lower band
which corresponds to a Mott-type metal-insulator transition.

However the existing experimental estimates of the activation energies
for $ \varepsilon_{2}$ nearest-neighbor hopping conductivity give
values much less than theoretical estimates (about 10 meV) - as small as
to say 1 meV and even smaller (see e.g. \cite{Polyan} and references
given in this paper and also \cite{Gersh}). The most reliable of these
estimates were based on studies of magnetoresistance which give a unique
possibility to reveal the $ D^{-}$ band contribution. According to our
analysis of existing experimental data \cite{Agrin}, not only nearly
non-compensated samples, but a great deal of experimental data for
nearest neighbor hopping ( with $K=0.4-0.7$) show a signature of $D^{-}$
band contribution.

Here we are not aimed to discuss in detail a problem why the Hubbard
energy can be as small. The polaron-type effects we refer can be an
explanation. One has to keep in mind that not only small 
Hubbard energies, but
even negative (and large!) Hubbard energies are possible for many centers
(to say for {\it Dx} -centers). Then, the Hubbard energy is known to be
suppressed near the metal-insulator transition due to a divergency of
both localization length and dielectric constant, the suppression being
more effective than a suppression of a width of the impurity band.

We would like also to note that in the $2D$ structures, the number of
electrons in the 2D layer at the dielectric side of MIT can be even
larger than the total number of localized centers within this layer
since, in contrast to 3D samples, the neutrality condition holds only for
the whole structure including remote dopants far away from the
layer 
(note that this fact was noted by Klapwijk and Das Sarma  \cite{Klap}
This latter fact in a natural way favors a formation of $D^{-}$ centers.

First let us recall the scenario of occupation of the $D^{-}$ band
considered earlier in \cite{Agrin}. According to these estimates the
doubly occupied centers have a distribution function 
\begin{equation}\label{D-}
n_{D^{-}}= \frac{1}{\exp (\frac{2\varepsilon +U-2\mu }{T})+2\exp (\frac{
\varepsilon -\mu +U}{T}) \cosh (\frac{\mu_{0}gH}{2T})+1}  
\end{equation} 
where $\mu $ is chemical potential; in the case when the
total number of electrons is dominated by single-occupied sites while the
total number of $ D^{-}$ sites is small 

\begin{equation} 
\mu =\mu
(H=0)-T \ln (\cosh \frac{\mu _{0}gH}{2T}).  
\label{muH1} 
\end{equation}

As it is seen, the choice between the two exponentials in the denominator
depends on sign of $\varepsilon -\mu $: for $\varepsilon -\mu >0$ the
first one dominates, while for $\varepsilon -\mu <0$ the second one
dominates. However, the magnetic field contribution is the same to both
of the exponent since the factor $2\cosh (..)$ is actually an addition to
the exponent equal to $\mu -\mu (H=0)$.

Let us show that the presence of $D^-$ band can lead to a pronounced
positive spin magnetoresistance for hopping conductivity.

One expects the $D^{-}$ states to be formed in tail of the standard
impurity band. Thus let us discuss in more detail the features of
band-tail hopping. Let us assume that the tail can be described by an
exponential decay 
\begin{equation} 
\nu (\varepsilon )=\nu _{0}\exp
(\frac{\varepsilon -\varepsilon _{0}}{ \varepsilon _{1}}) 
\end{equation}

where $\varepsilon _{0}$, $\varepsilon _{1}$ are some constants; $
\varepsilon <\varepsilon _{0}$. Actually for $D^{-}$ band the position of
the Fermi level is by the value $U$ lower than for single-occupied states
(so in the distribution given by Eq. (\ref{D-}) one should take
$\varepsilon \rightarrow -\varepsilon -U$). Let us first consider a
situation with no Coulomb gap. If the energy $\varepsilon _{1}$
characterizing a decay of the bandtail is larger than the typical hopping
energy band, the hopping is of standard Mott type. However, if the
hopping band is larger than $\varepsilon _{1}$, the situation is
different. Indeed, paying a higher activation energy one is able to find
a closer hopping site (then typical for standard VRH) because of a strong
density of states increase with energy increase. Since the hopping length
corresponding to site energy $\varepsilon $ is \[ r\sim
(n_{0}(\varepsilon )\varepsilon )^{-1/3} \] one can find $\varepsilon $
in a standard way comparing tunneling and activation contribution to the
hopping exponent: 
\begin{equation} 
\lbrack \nu (\mu )\exp
(\frac{\varepsilon -\mu }{\varepsilon _{1}} )(\varepsilon -\mu
)]^{1/3}(\varepsilon -\mu )\sim \frac{2T}{a} 
\end{equation} 
and thus
\begin{equation} 
\frac{\varepsilon -\mu }{3\varepsilon _{1}}\sim \ln
\frac{T}{(\varepsilon -\mu )^{4/3}a\nu (\mu )^{1/3}} 
\end{equation} 
Thus
one has for the conductivity

\begin{equation}\label{log} 
\sigma \propto \exp
(-\frac{\varepsilon -\mu }{T})=\exp (-\frac{3\varepsilon _{1}\ln
[T/(\varepsilon _{0}-\mu )^{4/3}a\nu (\varepsilon
_{0})^{1/3}]+(\varepsilon _{0}-\mu )}{T}).  
\end{equation}

As it is clearly seen, the tail hopping has a character of activated
behavior (if logarithmic term is considered as a constant).

Certainly one would have a purely Arrenius law if the $D^{-}$ band would
have a sharp edge and the chemical potential would cross this edge after
the application of a strong magnetic field. It is the bandtail states
with exponential decrease of density of states that ensure a presence of
the logarithm term. As it is seen, the latter makes the temperature
behavior to be weaker than the Arrenius law. Note that if we would assume
a power decrease of DOS in the tail ($\propto \varepsilon ^{-\alpha }$,
$\alpha >>1)$ we would obtain a power law addition in the numerator of
the exponent in Eq. (\ref{log}) of the sort $\sim -\varepsilon
_{1}[(\varepsilon _{0}-\mu )^{4/3}an(\varepsilon _{0})^{1/3}/T]^{1/\alpha
}$. One notes that in our situation the activation is to the states
corresponding to the boundary of the tail region which is characterized
by $\varepsilon _{0}$.

If one includes the Coulomb gap into considerations, it would lead to a
quadratic gap near the Fermi level, but the strong energy dependence of
"bare" density of states does not allow for this gap to develop up to its
nominal value according to the value of $n(\mu)$ due to a cut-off at
energies $\sim \varepsilon_1$ controlling the DOS decay.

Now let us turn to the magnetic field dependence which is the most
important for us. As it is clearly seen, the main effect is a dramatic
decrease of DOS at Fermi level. Thus one expects that if at $H=0$ the DOS
at the Fermi level is large enough and the Coulomb gap hopping can be
observed, the gradual deepening of Fermi level with $H$ increase would
change the situation to the activation-like behavior with no significant
region for Mott-type hopping.

\section{Experiment}

The sample investigated here is a delta-doped GaAs/AlGaAs heterostructure
which has been used in previous studies \cite{Trem,Khon}, where full
details of the layer composition, doping and device fabrication are
given. The low-temperature measurements of conductivity in strong
magnetic fields were carried out in the Cavendish Laboratory, University
of Cambridge. The longitudinal resistivities at different carrier
concentrations $n$, temperature $T$ and magnetic field $B$ were measured
from the Ohmic part of the dc four probe $I-V$ characteristics. The data
presented here correspond to field parallel to the 2D plane and current
being parallel to the field. We have also measured the case where the
magnetic field is parallel to the 2D field and perpendicular to the
current and no anisotropy was observed.

Figure 1 shows the logarithm of resistivity plotted versus $1/T$ for $
n=9.52\times 10^{10}cm^{-2}$ at $B=0$ (bottom), 6 (middle) and 8 Tesla
(top curve). At $B=0,$ the data follow $T^{-1/3}$ behavior for $T>1$ K
while it follows $T^{-1/2}$ behavior for $T<1$ K \cite{Khon}. However,
application of magnetic field causes a strong increase in resistivity
which is larger at lower temperatures. This causes a change in hopping
behavior. At $B=6$ and 8 T the data follow $T^{-1/2}$-law for $T>1$ K
and $T^{-0.8}$ behavior for $ T<1$. Similar results were obtained for
$n=9.84\times 10^{10}$ cm$^{-2}$ and $n=9.18\times 10^{10}$
cm$^{-2}$ for the magnetic field
regime of B=4, 6, 8 and 10 T although at lower
magnetic field the transition to the $T^{0.8}$ behavior
shifts towards lower temperature while at higher field
it shifts towards higher temperature.
For the limited interval of temperatures, one can approximate the
experimental curves by simple Arrenius law with constant energy of
activation $E$. 

On the Fig. 2 scaling of the magnetoresistance at different
temperatures  in terms of the ratio
$H/T$ is demonstrated for $n=9.18\times 10^{10}$ cm$^{-2}$ . 
It is seen that the slope is the same
for all the fields studied while a slight shift
of the curve with an increase of magnetic field is related to
(small) orbital contribution to magnetoresistance.

Figure 3 shows the activation energy $E$ plotted versus $B$ for different
carrier concentration. The slope of the straight lines correspond to
$g\mu _{B}$. From these slopes we calculate the values of $g$ as
$0.08,0.10$ and $ 0.115$ at $n=9.84,$ $9.52$ and $9.18\times 10^{10}$
cm$^{-2}$ respectively. The values of $g$ are substantially lower than
that of the bulk GaAs value ($ g=0.44$) and are decreasing with
increasing $n$.

\section{Discussion}

Now we would like to compare theoretical predictions with the
experimental data. In the previous paper \cite{Shlimak}, the $T^{-0.8}$
behavior was interpreted as $T^{-1/2}$ law with the temperature-dependent
prefactor. This interpretation is based on the assumption that mechanism
of conductivity in strong magnetic fields is still VRH, but with
reconstruction of the phonon assistance. However, there is another
possible interpretation of the $ T^{-0.8} $ behavior: one can suggest
that it is slightly corrected Arrenius law ($T^{-1}$) as it is shown on
Fig. 1. This means that the mechanism of conductivity in strong fields is
changed and is determined by excitation of localized carriers to the
states in the upper Hubbard band with larger radius of localization. In
the present paper we investigate this possibility.

As it is seen, there is at least qualitative agreement between
theoretical model discussed above and the experiment. Indeed, the
experiment exhibits a transition from Efros-Shklovskii type of hopping to
nearly Arrenius law (with an exponent $(T^{\prime}/T)^{0.8}$) when a
strong magnetic field is applied. The difference between this behavior
and the Arrenius law can be explained by the finite density of states in
the bandtail according to a scenario discussed above.

The magnetic field behavior of the activation energy is close to linear
law in strong field limit. The value of the effective $g$-factor
extracted from a comparison of experimental curves with the strong field
limit of Eq. (\ref {muH1}) gives a value about 4 times lower than the
handbook values for GaAs. However, one has in mind that we deal with
AlGaAs-GaAs heterostructure rather than with the bulk GaAs. The
$g$-factor values for AlGaAs quantum wells were calculated theoretically
\cite{Ivchenko}. It was shown that due to the fact that in such
structures one has a mixture of GaAs states (for which the $g$-factor
$\sim -0.45$ is negative) and AlGaAs states (where $g$ -factor is
positive) the effective $g$-factor depends on the well width $d$ and can
even vanish for $d\sim 5$ nm. Such theoretical predictions agree with
experimental data (see e.g. \cite{Gravier}). Since we deal with a similar
system where the effective width is controlled by the gate voltage we
believe that these considerations can be applied. In particular, a
decrease of gate voltage makes the effective field in the structure to be
stronger which corresponds to a decrease of the well width (for the
electrons at the Fermi level) and leads to a decrease of $g$-factor. Fig.
2 clearly demonstrates such a decrease of $g$-factor with a decrease of
the gate voltage in agreement with above mentioned considerations.

Let us estimate quantitatively the variation of $g$-factor with the gate
voltage. First we estimate the derivative ${\rm d}g/{\rm d}F$ where $F$
is an effective electric field within the well. Using the values of
$g$-factors reported in \cite{Ivchenko} for biased GaAs-AlGaAs quantum
wells for two different biases, one could estimate ${\rm d}g/{\rm d}F\sim
0.2\cdot 10^{-5}$ V$^{-1}$cm which gives $\delta g\sim 0.02$ for the
experimental variation of $V_{g}$ equal to 0.01 V. Unfortunately, both
limited set of biases and finite well widths (in contrast to our case)
make this estimate to be a rough one. Another approach to estimate the
derivative in question may use the fact that the decrease of $g$-factor
with respect to its bulk value observed for our structures is related to
the effective electric field $F$; thus one can estimate ${\rm d}g/{\rm
d}F\sim (g-g_{bulk})/F$. Then, the variation of $F$ due to variation
$\Delta V$ of the gate potential can be estimated as $\Delta n/n\sim
\Delta F/F$ where $\Delta n$ is a variation of the carrier concentration.
Thus one has $\Delta F\sim (\Delta n/n)F$ and, consequently, $\Delta
g\sim (g-g_{bulk})(\Delta n/n)$ which for $g\sim 0.1$, $g_{bulk}=0.45$,
$\Delta n/n\sim 0.075$ one has $\Delta g\sim 0.025$. Since the variation
of $g$ observed is about $0.035$ one can conclude that the agreement with
our rough estimates is at least a reasonable one.

Thus we believe that both the values of $g$-factor extracted from the
magnetoresistance data and its voltage dependence give an evidence for
our theoretical model.

Since the system studied is rather close to ones exhibiting 2D MIT, we
find it of interest to consider a possible consequences of our model for
explanation of the details of this MIT.

Since our results imply a significant role of $D^{-}$ band for the
structures in question, we believe that due to much larger value of the
localization length, the conductance is controlled by $D^{-}$ states even
if they have smaller DOS at the Fermi level than single-occupied states.
In other words, we believe that MIT takes place in the upper Hubbard band
(when chemical potential crosses a mobility edge in $D^{-}$ band) and
thus has features of Mott transition.

Moreover, one can expect that for the lower Hubbard band the mobility
edge is situated above the chemical potential at some energy $\varepsilon
_{1,m},$ that is the corresponding states are localized. In this case we
expect the situation completely controlled by the value of the DOS at the
Fermi level of $D^{-}$ band given by Eq. (\ref{D-}) for $\varepsilon =\mu
-U$.

First, one notes that an external magnetic field can in this case lead to
a decrease of DOS due to a gradual lowering of the Fermi level in $D^{-}$
band according to the behavior of the chemical potential, Eq.
(\ref{muH1}). However we believe that the MIT takes place when DOS is
still high enough to support the metallic conductivity, and its energy
dependence near MIT is not as strong as in the scenario of activated
hopping discussed above.

In this case one can write for the change of the electron concentration
in $ D^-$ band 
\begin{equation}  \label{deltan} 
\Delta n (H) = n(H) -
n(0) = \nu (\mu(H=0))\delta \mu = - \nu T \ln (\cosh \frac{\mu_0 g
H}{2T}) 
\end{equation} 
Since the decrease of concentration near MIT leads
to a corresponding decrease of conductivity, one expects that at least
for small $\Delta n$ this decrease can be described as 
\begin{equation}
\Delta \sigma = \sigma (H,T) - \sigma (0,T) =\frac{\partial
\sigma}{\partial n} \delta n \equiv \sigma^{\prime}_n(T) \delta n(H,T)
\end{equation} 
Considering $\Delta \sigma (H,T)$ behavior one concludes
that $H-T$ scaling law is given by 
\[ \sigma^{\prime}_n(T)\nu T \ln
(\cosh \frac{\mu_0 g H}{2T}) = const \] 
and thus depends on the
temperature dependence of conductivity.

However, deep at the insulating side of MIT, when the activation to the
mobility edge in $D^{-}$ band dominates, the conductivity is controlled
by a corresponding activation exponent, the magnetic-field dependent
contribution to this exponent is equal to 
\begin{equation} \frac{\delta
\mu (H)}{T}=\ln (\cosh \frac{\mu _{0}gH}{2T}) 
\end{equation} 
which
depends only on a ratio $H/T$.

The paper \cite{Sar} reported a suppression of the metallic phase at
strong magnetic field followed by a saturation of resistance, the
behavior exhibited $H/T$ scaling while the most representative data
corresponded to the dielectric side of MIT. We would like also to mention
the paper \cite {Sar1} reporting large positive magnetoresistance deep in
the dielectric side of 2D MIT followed by the saturation plateau. We
believe that the arguments given above can explain such a behavior.
(Note that the qualitative arguments
relating magnetoresistance on the dielectric side of 2D MIT
to an elimination of the condition for electron pairs
binding were also given by
Klapwijk and Das Sarma in  \cite{Klap}).
We would like also to emphasize that the saturation behavior observed at
strong field limit at the insulating side of MIT in \cite{Sar,Sar1}, can
be easily explained in our model as a result of a competition between
suppressed $D^{-}$ band contribution and a contribution of lower Hubbard
band not affected by magnetic field (a similar behavior was predicted in
\cite{Agrin} for nearest-neighbor hopping). Note that a decrease of
resistance at saturation observed in \cite{Sar1} with an increase of
electron concentration can be explained as a result of an increase of
conductivity supported by the lower Hubbard band. Another important
feature reported in \cite{Sar1} is a fact that while for dielectric side
of MIT the magnetoresistance is pronounced already at weak fields, for
the metallic side there exists a region of weak fields with no
significant magnetoresistance, and the width of this region increases
with increase of concentration. This behavior is in agreement with our
model where the pronounced magnetoresistance corresponds to a position of
the chemical potential within the band tail of $D^{-}$ band while at
metallic state initial position of the chemical potential corresponds to
high enough DOS.

Now let us discuss the temperature behavior of conductance in our model
for the metallic side of MIT. One sees that while at $T$ = 0 all the
single-occupied states are localized, at elevated temperatures an
activation takes place to the mobility edge within the bandtail
of the 2D conductance
band. As a result, the empty (positively charged) centers are created
which are effective scatterers for the conducting electrons of the
$D^{-}$ band. Thus one expects that an increase of temperature is
accompanied by an increase of resistance.

The corresponding increase of resistivity is obviously proportional to a
number of empty states under the mobility edge. Thus at small
temperatures the resistance is expected to follow the law
\begin{equation} 
\rho (T)=\rho _{0}+\rho _{1}\exp (-T_{s}/T)  \label{res}
\end{equation} 
where $T_{s}=\varepsilon _{m,1}-\mu (T=0).$ If the
effective width of the single-occupied band is of the order or less than
$\varepsilon _{m,1}-\mu (T=0),$ one expect that the resistance is
saturated when $T>>T_{s}$ that is the law Eq. (\ref{res}) holds
approximately for all temperatures. The exact temperature behavior should
take into account the temperature dependence of chemical potential
calculated for the given details of the band density of states.

One also notes that if there is some gap between the band of localized
states and 2D conductance band the scenario discussed above holds even if
nearly all of the localized states are occupied that is when the position
of the Fermi level within the lower Hubbard band is high and the initial
number of charged scatterers is small. In this case the situation $\rho
_{0}<<\rho _{1}$ (observed, in particular, in \cite{Krav}) can be
realized.

Note that the temperature behavior of resistance can be also affected by
a variation of a number of electrons in the $D^{-}$ band with temperature
increase due to electrons exchange between the two Hubbard bands which
would correspond to temperature dependencies of both $\rho _{0}$ and
$\rho _{1}$. However we would like to note that we assume that the MIT
takes place not at the ''bandtail'' of $D^{-}$ band, but when the Fermi
level is far in the $ D^{-}$ band and the behavior of density of states
in the $D^{-}$ band is not strong. Thus if the number of electrons in
$D^{-}$ band increases with temperature, it can lead to weak decrease of
resistance with temperature increase which is indeed observed after
initial (exponential) growth.

We would like to note that the latter mechanism of resistance in the
metallic state can be considered as a modification of the mechanism
suggested earlier by Altshuler and Maslov \cite{Altshuler} to describe
the experimental data in question. The difference is that Altshuler and
Maslov have related the scattering centers with some traps with energies
close to the Fermi level situated within the insulating layer of the
MOSFET's. At the same time Altshuler and Maslov have also considered a
possibility of double occupation of the trap leading to a spin mechanism
of magnetoresistance due to effect of magnetic field on the charge states
of the scatterers.

In our picture we consider MIT within the upper Hubbard band and relate
the scattering centers to the single-occupied states of the lower Hubbard
band. In contrast to Altshuler and Maslov, the magnetoresistance in our
case originates due to magnetic-field driven change of a position of
$D^-$ band with respect to the Fermi level and thus is related to change
of carrier concentration rather than to a change in the scatterer system.
In our model the latter is expected to be related to the lower Hubbard
band states, their occupation numbers being independent of magnetic
field.

For our scenario, a presence of scattering centers able to change their
charge states at low temperatures is an inherent property of the model
and does not need any additional assumption. Then, our model can also
explain features of the 2D MIT observed in gated GaAs heterostructures
\cite{heter} where the significant concentration of traps is not
expected.

The latter devices seem to be similar to devices studied in this paper
experimentally. We believe that these facts give evidence to support our
model. Indeed, it describes features observed for the devices in question
both in the dielectric limit (Ref. \cite{Shlimak} and this paper) and
near the MIT, Ref. \cite{heter}.

Now let us discuss in some more detail a possible realization of the
scenario discussed above in our Si $\delta $-doped GaAs-AlGaAs
heterostructure. If there would be no delta-dopants, the application of
the gate potential depleting the 2DEG would lead to localization of
electrons in the large-scale potential created by doping impurities
situated beyond the spacer. The delicate feature is that the parameters
of this potential depend on screening and thus on the electron
concentration in the quantum well. The delta-dopants, which act as strong
Coulomb centers situated nearly in the 2D conducting layer, apparently
manifest itself as ''embryos'' for localization. Namely, we can expect
that the localization in course of depletion would start just from
filling these states. If so, we can expect that at certain gate potential
one would have electrons localized at delta-dopants and, in addition,
some delocalized electrons or electrons localized in the rest of
large-scale potential. The system of the delta-dopants in this case could
be considered as a completely occupied impurity band with energies
situated some lower than to say energies of the large-scale potential. We
can speculate that just the delta-dopant states form the $D^{-}$ band.

To conclude, we have suggested a model allowing to explain unusual
features of hopping conductivity in parallel magnetic fields and
metal-insulator transition recently reported for 2D structures. The model
implying a dominant role of the upper Hubbard band explains in a natural
way both the suppression of the metal state by the magnetic field with
$B/T$ scaling and magnetic-field driven crossover from VRH to activated
hopping as a result of the on-site spin correlations. . The dramatic
increase of resistance with temperature increase observed in metallic
state is ascribed to activation of electrons from initially localized
single-occupied states to the mobility edge in the tail of 2D
conductivity band leading to a creation of additional scattering centers
in a way similar to the one considered by Altshuler and Maslov
\cite{Altshuler}. The experimental studies of the hopping
magnetoresistance for Si $\delta $-doped GaAs-AlGaAs heterostructure give
additional evidence for the model.

\section{Acknowledgements}

NVA and VIK are grateful to E. L. Ivchenko for valuable discussions and
acknowledge a financial support by Russian Foundation for Fundamental
Research. SIK and IS are thankful to Professor M. Pepper, J. T. Nicholls
and D. A. Ritchie from Cavendish Laboratory, University of Cambridge for
significant help in the low-temperature measurements.

\vskip3cm {\large Figure captions}

\medskip

Fig.1 Resistivity $\rho $ of Si $\delta $-doped GaAs-AlGaAs
heterostructure plotted as $\ln \rho $ versus 1/$T$ for $B$=0, 6, 8 Tesla
(from bottom to top) at $n=9.52\times 10^{10}$ cm$^{-2}$.

Fig.2 Magnetoresistance as a function of $H/T$ for $n =
9.18\cdot 10^{10} cm^{-2}$.

Fig.3 Activation energy of conductivity $E$ plotted versus $B$ for
electron densities $n=$ 9.18, 9.52 and 9.84$\times 10^{10}$cm$^{-2}$
(from {\it 1} to {\it 3}).

\end{document}